\documentclass[journal]{IEEEtran}

\ifCLASSINFOpdf
\else
\fi

\usepackage{graphicx}
\usepackage{amsmath}
\usepackage{cite}

\begin{document}
\title{Open Set Modulation Recognition Based on Dual-Channel LSTM Model}

\author{Youwei~Guo,
        Hongyu~Jiang,
        Jing~Wu,
        and~Jie~Zhou%
\thanks{Y. Guo, H. Jiang, J. Wu, and J. Zhou are with the Institute of Electronic 
Engineering, China Academy of Engineering Physics, Mianyang 621999, China 
(email:pkugyw@pku.edu.cn;doherty2004@163.com;wu1019@mail.ustc.edu.c-
n;zhoujie-iee@caep.cn).}}%

\maketitle

%
\IEEEpeerreviewmaketitle

\begin{abstract}
Deep neural networks have achieved great success in computer vision, speech recognition and many other areas. The potential 
of recurrent neural networks especially the Long Short-Term Memory (LSTM) for open set communication signal modulation recognition 
is investigated in this letter. Time-domain sampled signals are first converted to two normalized matrices which will be fed into a 
four layer Dual-Channel LSTM network tailored for open set modulation recognition. With two cascaded Dual-Channel 
LSTM layers, the designed network can automatically learn sequence-correlated features from the raw data. With center loss and 
weibull distribution, proposed algorithm can recognize partial open set modulations. Experiments on the public RadioML dataset 
indicates that different analog and digital modulations can be effectively classified by the proposed model, while partial 
open set modulations can be recognized. Quantitative analysis on the dataset shows that the proposed method can achieve an 
average accuracy of 90.2\% at varying SNR ranging from 0dB to 18dB in classifying the considered 11 classes, while accuracy 
of open set experiment dramatically improved by 14.2\%.  
\end{abstract}

\begin{IEEEkeywords}
Deep learning, DC-LSTM, LSTM, modulation recognition, open set.
\end{IEEEkeywords}

\section{Introduction}
\IEEEPARstart{T}{he} recognition of modulation signals plays an important role in cognitive radio and spectrum monitoring owing to 
the rich information contained in communication signals. Numerous automatic modulation recognition algorithms have been proposed 
which mainly include decision theory based methods and machine learning based methods 
\cite{dobre2007survey}.%
\par{}\setlength{\parindent}{2em}The letter only focuses on machine learning based algorithms due to the huge demand of parameters 
estimation of decision theory based algorithms. Current machine learning based algorithms can be divided into two types: methods 
based on traditional machine learning and methods based on deep learning. Traditional machine learning based modulation recognition 
algorithms need to design handcrafted expert features extractor and combine features with machine learning algorithms appropriately. 
Deep learning has achieved great success in computer vision and many other areas \cite{lecun2015deep}. They have also been applied 
to modulation recognition. Researchers have utilized deep neural network to improve the performance of classifiers 
\cite{fu2015deep}, to recognize modulations by constellation diagram \cite{zhu2017modulation}, 
by feature graph \cite{peng2017modulation,zhang2017modulation} or by sampling signals \cite{o2016convolutional,west2017deep,
rajendran2017distributed}.%
\par{}\setlength{\parindent}{2em}
Nevertheless, much works so far just only focus on modulation classification of close set, hence open set unknown classes cannot 
be rejected. Not only the deep neural networks can be used to extract features automatically from modulation signals, they can also 
be used to recognize open set signals. Open set deep neural network based on weibull distribution is proposed in 
\cite{bendale2016towards} to recognize images of unknown classes, however the activation vectors of which are separable instead 
of discriminative. The Dual-Channel LSTM network (DC-LSTM) which achieves state of the art classification performance is firstly 
introduced, then DC-LSTM will be combined with center loss \cite{wen2016discriminative} and weibull distribution 
\cite{bendale2016towards} to realize open set modulation recognition.%
\par{}\setlength{\parindent}{2em}The proposed scheme consists of the following steps:1) convert the complex communication 
modulation signal to two matrices which are appropriate as a recurrent neural network input;2) two cascaded 
Dual-Channel long short-term memory layers designed to extract sequence-correlated features of 
In-phase/Quadrature signal components and Amplitude/Phase signal components;3) two fully connected layers designed to concatenate 
features and classify features to corresponding classes;4) An extreme value distribution - weibull distribution adopted to fit 
the cut off probability of the distance from features to features centers and modify neural network outputs. The effectiveness 
of the methodology is validated on a standard public dataset RadioML2016.10a \cite{o2016radio}. Quantitative analysis indicated 
that the proposed method can achieve the state of the art performance and recognize open set modulations. It is the first effort 
to achieve open set modulation recognition to our best knowledge.%
\section{Methodology}
\subsection{Two Real Matrices Representation for Modulation Signal}
The general representation of the received signal can be fully expressed as a complex vector which is given by:
\begin{equation}\label{eq1}r(n) = s(n)*c(n)+N(n)\end{equation}
\par{}\setlength{\parindent}{0em}where $s(n)$ is the emitting noise free complex baseband envolop of the received signal $r(n)$, 
$N(n)$ is Additive White Gaussian Noise (AWGN) with zero mean and variance $\sigma^2_n$ and $c(n)$ is the time varying impulse 
response of wireless transmission channel \cite{rajendran2017distributed}. It has to be converted to real matrices in order to 
be able to be fed into real-valued recurrent neural networks. Two matrices representation $V_1$,$V_2$ tailored for neural 
networks are proposed as follows:
\begin{equation}\label{eq2}R  = \sqrt{\frac{1}{T}\sum_{n=1}^T{|r(n)|}^2}\end{equation}
\begin{equation}\label{eq3}I(n)  =\Re(r(n))/R\end{equation}
\begin{equation}\label{eq4}Q(n)  = \Im(r(n))/R\end{equation}
\begin{equation}\label{eq5}A(n)  = \sqrt{I^2(n)+Q^2(n)}/R\end{equation}
\begin{equation}\label{eq6}P(n)  = arctan\left(Q(n)/I(n)\right)/{\pi}\end{equation}
\begin{equation}\label{eq7}V_1 = [I,Q]\end{equation}
\begin{equation}\label{eq8}V_2 = [A,P]\end{equation}
where $T$ is the sample length of the received signal $r(n)$, $I(n)$, $Q(n)$ and $A(n)$ are the real part, imaginary part 
and the instantaneous amplitude of the received signal $r(n)$ which are L2-Normalized by $R$, $P(n)$ is the instantaneous 
phase of the received signal $r(n)$ which is normalized between -1 and +1.%
\subsection{Dual-Channel LSTM}
\begin{figure}[!t]
  \centering
  \includegraphics[width=3in]{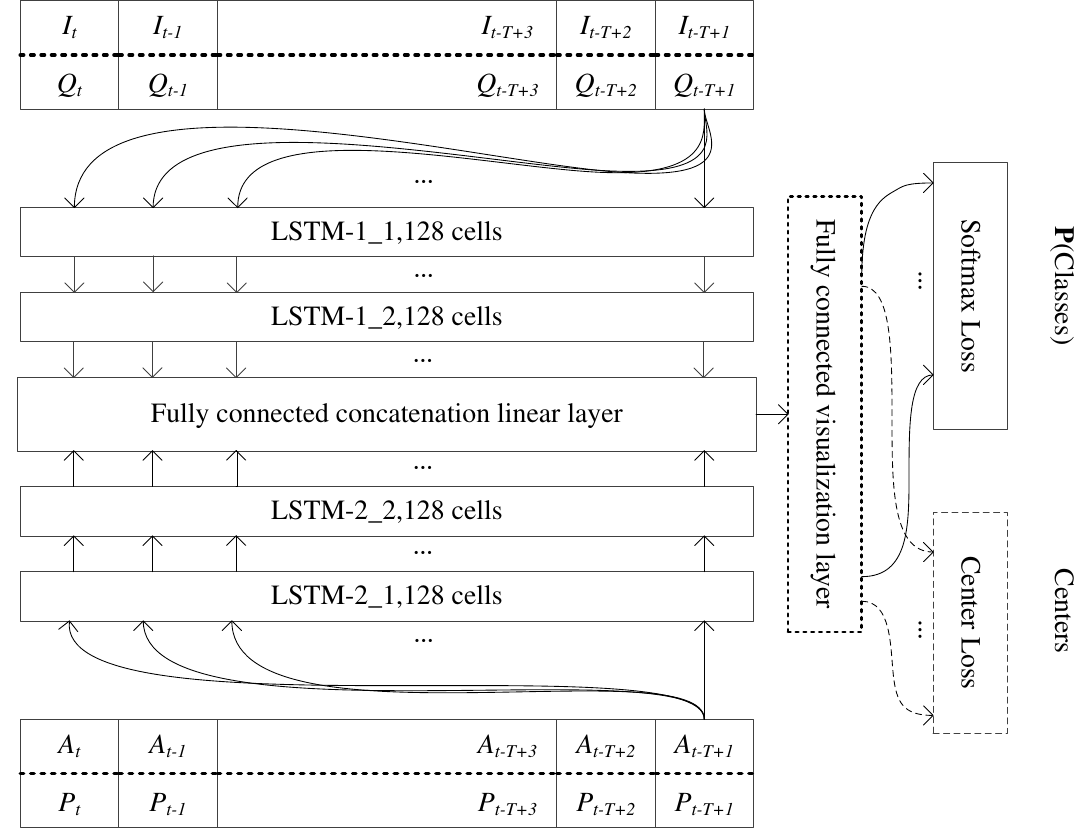}
  \caption{Dual-Channel LSTM network architecture.}
  \label{DCLSTM}
\end{figure}
\par{}\setlength{\parindent}{2em}The architecture of Dual-Channel LSTM (DC-LSTM) in this letter is shown in Fig. \ref{DCLSTM} which 
contains two Dual-Channel LSTM layers, two fully connected layers, a softmax classifier and a centers calculator which is connected 
to the penultimate layer output.%
\par{}\setlength{\parindent}{2em}Each LSTM layer is comprised of 128 LSTM cells and the activation function is hyperbolic tangent. 
Different LSTM layer channels are designed to extract features of different attributes(such as I/Q or A/P). Detailed parameters 
of each layer are descibed as follows.%
\par{}\setlength{\parindent}{2em}Dimension of each training or testing example is $C$$\times $$T$$\times $$2$, where $C$ represents 
the channel of input signal and $T$$\times $$2$ represents the dimension of signal components. Each time step representations 
with dimension $C$$\times $$2$ are fed into the network, an output of $C$$\times $$T$$\times $$C_1$ obtained after $T$ time steps, 
where $C_i$ represents the cell number of $i_{th}$ LSTM layer, for the second LSTM layer only the last time step will produce an 
output whose dimension is $C$$\times $$C_2$, the output of different channels are concatenated by concatenation layer and then 
will be fed to the fully connected layers. A fully connected layer with two neurons will be connected to the concatenaion 
layer for visualization, and the output layer has $N$ (number of modulations classes) neurons.%
\subsection{Loss Function}
\par{}\setlength{\parindent}{2em}The loss function of proposed network is different from traditional classification neural networks 
whose loss funciton only contains cross-entropy (also named Softmax Loss). A loss function which combines cross-entropy with center 
loss is introduced in \cite{wen2016discriminative}, with which minimal classification error and clustering features can be obtained 
simultaneously. It can be written as:
\begin{equation}
  \label{eq9}
  \begin{aligned}
    L = L_s+\lambda{L_c} = &-\sum_{i=1}^mlog\frac{e^{{W_{y_i}^Tx_i}+b_{y_i}}}{\sum_{j=1}^n{e^{{W_{y_j}^Tx_i}+b_{y_j}}}}+\\
  &\frac{\lambda}{2}\sum_{i=1}^m\|x_i-c_{y_i}\|_2^2 
\end{aligned}
\end{equation}
where $x_i$ represents the $i_{th}$ output of the penultimate layer, $L_s$ and $L_c$ represent cross-entropy and center loss 
respectively, $c_{y_i}$ represents the center vector whose label equal to $y_i$, $\lambda$ is the control parameter which 
balances cross-entropy and center loss.%
\subsection{Training and Open Set Testing}
\par{}\setlength{\parindent}{2em}Parameters needed to be updated of proposed model contain parameters of network architecture 
and classes centers. Parameters of network architecture are updated by Adam gradient descent algorithm 
with mini-batch size 256 and fixed learning rate 0.01. Centers are updated iteratively each mini-batch with initial centers 
equal to zero, $\lambda$ is set to 0.1 and  $\alpha$ is set to 0.5, detailed process can be found in \cite{wen2016discriminative}. 
Xaiver initial method is adopted to maintain the performance. All models are running on one workstation who has 128GB memory, 
K40c GPU and Xeon ES-2640 CPU with deep learning library Keras and Tensorflow.%
\par{}\setlength{\parindent}{2em}The special extreme value distribution - weibull distribution is adopted to fit the probability 
distribution of the distance from features to feature centers after classification model and feature centers acquired. The 
number M (which is determined by experience) is decided to choose how many farthest examples to fit the probability distribution 
funciton which is expressed as follows:
\begin{equation}
\label{eq10}
f(x|a,b) = \frac{b}{a}{\left({\frac{x}{a}}\right)}^{b-1}e^{-{\left({\frac{x}{a}}\right)}^b}I_{(0,\infty)}(x)
\end{equation}
where $a$ and $b$ control the scale and shape of the distribution, $I$ represents indicative function. Hence the cumulative 
distribution function can be written as:
\begin{equation}
\label{eq11}
F(x|a,b) = 1-e^{-{\left({\frac{x}{a}}\right)}^b}I_{(0,\infty)}(x)
\end{equation}  
the inverse cumulative distribution function is used to evaluate the probability of $x$ bigger than a value:
\begin{equation}
\label{eq12}
InvF(x|a,b) = 1 - F(x|a,b) = e^{-{\left({\frac{x}{a}}\right)}^b}I_{(0,\infty)}(x)
\end{equation}
the predictions can be modified by $a_i$,$b_i$ which are calculated from features $v^{(l-1)}$ and the corresponding centers $c_i$ 
as follows:
\begin{equation}
  \label{eq13}
  \hat{v}^{(l)}(i) = \left\{
             \begin{array}{lcl}
             {InvF\left(\|v^{(l-1)}-c_i\|^2|a_i,b_i\right)v^{(l)}(i)}&v^{(l)}(i)>0 \\
             {v^{(l)}(i)}&v^{(l)}(i)<0  
             \end{array}  
        \right.
\end{equation}
\begin{equation}
\label{eq14}
\hat{v}^{(l)}\left(N+1\right)=\sum_{i=1}^Nv^{(l)}(i)-\sum_{i=1}^N\hat{v}^{(l)}(i)
\end{equation}
\begin{equation}
\label{eq15}
\hat{P}(j)=\frac{exp\left(\hat{v}^{(l)}(j)\right)}{\sum_{i=1}^{N+1}exp\left(\hat{v}^{(l)}(i)\right)},j=1,2,...,N+1
\end{equation}
where $\hat{v}^{(l)}(N+1)$ and $\hat{P}(N+1)$ represent the activation value and probability of unknown class respectively. 
Besides, the mean accuracy metric is adopted for all experiments to evaluate the performance.%

\section{Experiments}
To evaluate the proposed framework, a standard public modulation signal dataset RadioML is considered, detailed parameters is 
shown in Table \ref{table1}. Two experiments - close set experiment and open set experiment are considered, where the 
training and testing set include all types in close set experiment and only testing set includes all types in open set experiment. 
These two experiments are used to evaluate the close set recognition performance and open set recognition performance respectively.
\newcommand{\tabincell}[2]{\begin{tabular}{@{}#1@{}}#2\end{tabular}}
  \begin{table}[!t]
   \centering
  \caption{\label{table1}RadioML2016.10a dataset parameters}
   \begin{tabular}{lcl}
    \hline\hline
   \tabincell{c}{Modulations} & \tabincell{c}{8PSK,AM-DSB,AM-SSB,BPSK,\\CPFSK,GFSK,PAM4,QAM16,\\QAM64,QPSK,WBFM} \\
    \hline
  Samples per symbol & 4 \\
  Sample length & 128 \\
  SNR Range & -20dB to +18dB \\
  Number of training samples & 110,000 \\
  Number of testing samples & 110,000 \\
    \hline
   \end{tabular}
  \end{table}

\subsection{Close Set Experiment}
Four LSTM based architectures - LSTM with IQ (LSTM-IQ), LSTM with AP (LSTM-AP), DC-LSTM, and DC-Bidirectional-LSTM 
(DC-BLSTM) are compared to choose the appropriate model. These models have been trained and tested on RadioML dataset and 
the training process is similar to \cite{rajendran2017distributed}. Parameters of different architectures, 
parameters number, and training time are listed in Table \ref{table2}.%
\par{}\setlength{\parindent}{2em}

After training 70 epochs on the dataset, DC-LSTM model can achieve an average accuracy of 90.2\% at varying SNR ranging from 0dB 
to 18dB. From the simulation results illustrated in Tabel. \ref{table2}, it can be concluded that whether utilizing IQ 
information or AP information, LSTM based model could achieve high recognition performance. This is significantly different from 
the results of \cite{rajendran2017distributed} where IQ information cannot be used to recognize modulations, it is mainly caused 
by the truth that IQ information is not normalized. In addition, with IQ and AP information combined, the performance of DC-LSTM 
achieves 2.2\% improvement than LSTM-AP model which has achieved the state of the art performace. DC-LSTM model is chosen as 
the network architecture for the following experiment considering parameters number, training time and classification 
performance. The results also indicate that  information in In-phase and Quadrature is more important than information in 
Amplitude and Phase. Besides, it is hard to declare which model is best among LSTM-IQ, DC-LSTM and DC-BLSTM.%
\renewcommand{\tabincell}[2]{\begin{tabular}{@{}#1@{}}#2\end{tabular}}
  \begin{table}[!t]
   \centering
  \caption{\label{table2}LSTM Based Modulation Recognition Architectures Comparison}
   \begin{tabular}{lclclcl}
    \hline\hline
   \tabincell{c}{Architecture} & \tabincell{c}{Cells per \\ layer} & \tabincell{c}{Parameters \\ Number} & \tabincell{c}{Training \\ Time} & \tabincell{c}{Accuracy \\ ($>=$0dB)}\\
    \hline
    AP 2 LSTM+1 Dense & 128 & 200,075 & 1.5h & 0.8790\\
    IQ 2 LSTM+1 Dense & 128 & 200,075 & 1.5h  & 0.9006\\
    DC 2 LSTM+1 Dense & 128 & 400,139 & 2.9h & 0.9023\\
    DC 2 B-LSTM+1 Dense & 128 & 1,062,411 & 9.4h & 0.8977 \\
    \hline
   \end{tabular}
  \end{table}
\subsection{Open Set Experiment}
\begin{figure}[!t]
  \centering
  \includegraphics[width=3in]{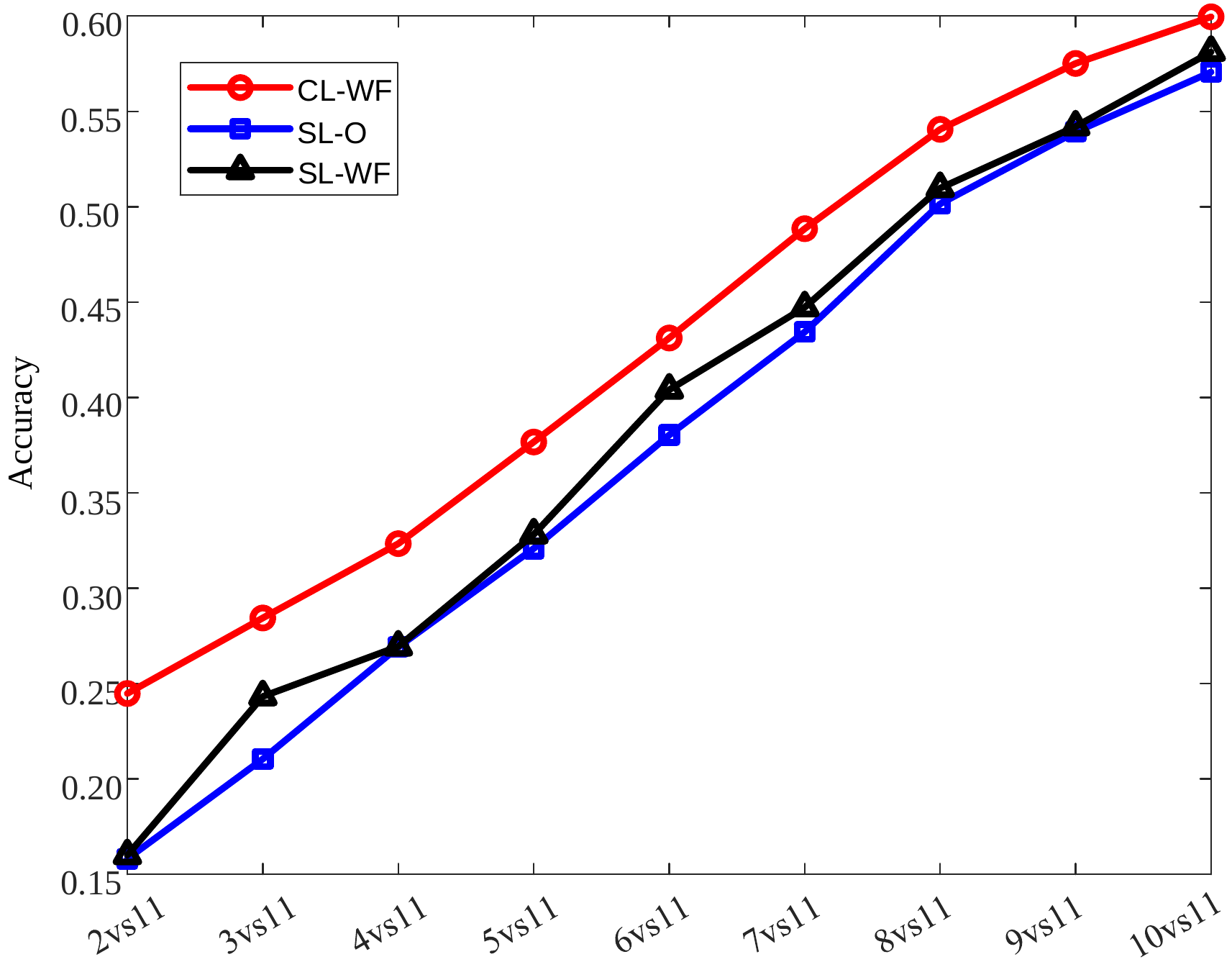}
  \caption{Open Set Performance Comparison.}
  \label{OSPC}
\end{figure}
To evaluate the open set recognition performance of the proposed model, the RadioML dataset is splitted into two subsets - training 
set and testing set where training set only contains partial types while testing set contains all types. Network weights and 
feature centers are acquired from training on the training set and weibull distribution parameters are calculated from centers and 
feature distribution. M is set to 1000 according to simulation experiments.%
\par{}\setlength{\parindent}{2em}Softmax loss only based model, softmax loss based model with weibull distribution fitting and 
center loss based model with weibull distribution fitting for open set recognition are shorted as SL-O,SL-WF and CL-WF for 
convenience. Comparison of different open set scene is illustrated in Fig \ref{OSPC} where “$a$ vs $b$” indicates that there are 
$a$ classes in training set and $b$ classes in testing set respectively. Test on the testing dataset reveals that proposed 
CL-WF model could achieve a significant improvement of 14.2\% than SL-O model. It should be noticed that error classified 
example in SL-O model which are ultimately modified by CL-WF model to ‘$unknown$’ class would be treated as rightly 
classified. From Fig \ref{OSPC} we can also see that proposed algorithm could recognize partial open set modulation types 
even the model has never seen before. Besides CL-WF model distinctly performs better than SL-WF model \cite{bendale2016towards}. 
The essential reason is that features in this model are not only separable but also discriminative.%
\par{}\setlength{\parindent}{2em}From section above we have seen that SL-WF model cannot tackle the recognition task when 
encountering unknown open set classes it has never seen while CL-WF model can recognize partial unknown classes. The following 
paragraphs give out the reason why the proposed model works by visualizing the features output of CL-WF model.%
\par
\begin{figure}[!t]
  \centering
  \includegraphics[width=3in]{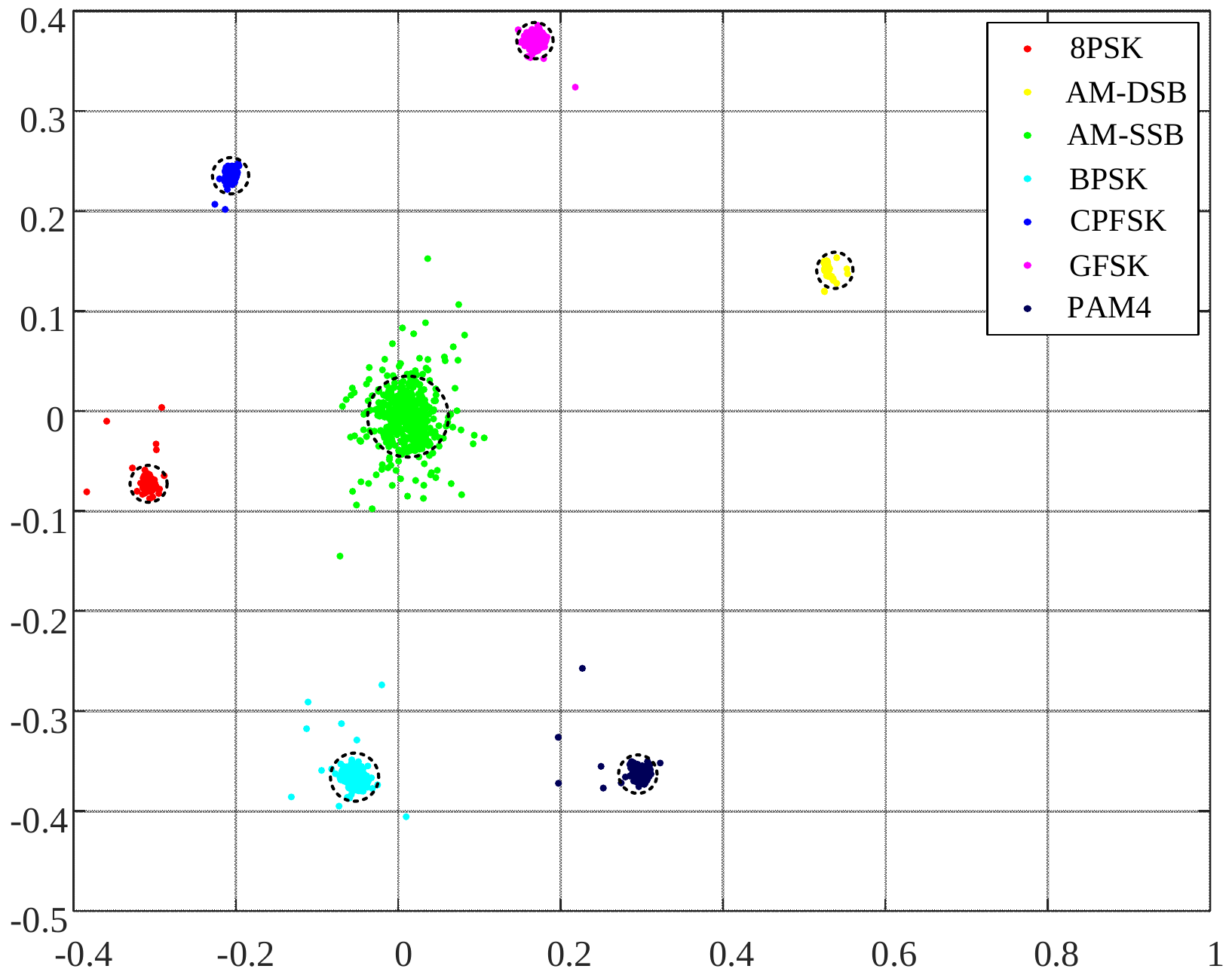}
  \caption{Features distribution of training set.}
  \label{trainft}
\end{figure}

\begin{figure}[!t]
  \centering
  \includegraphics[width=3in]{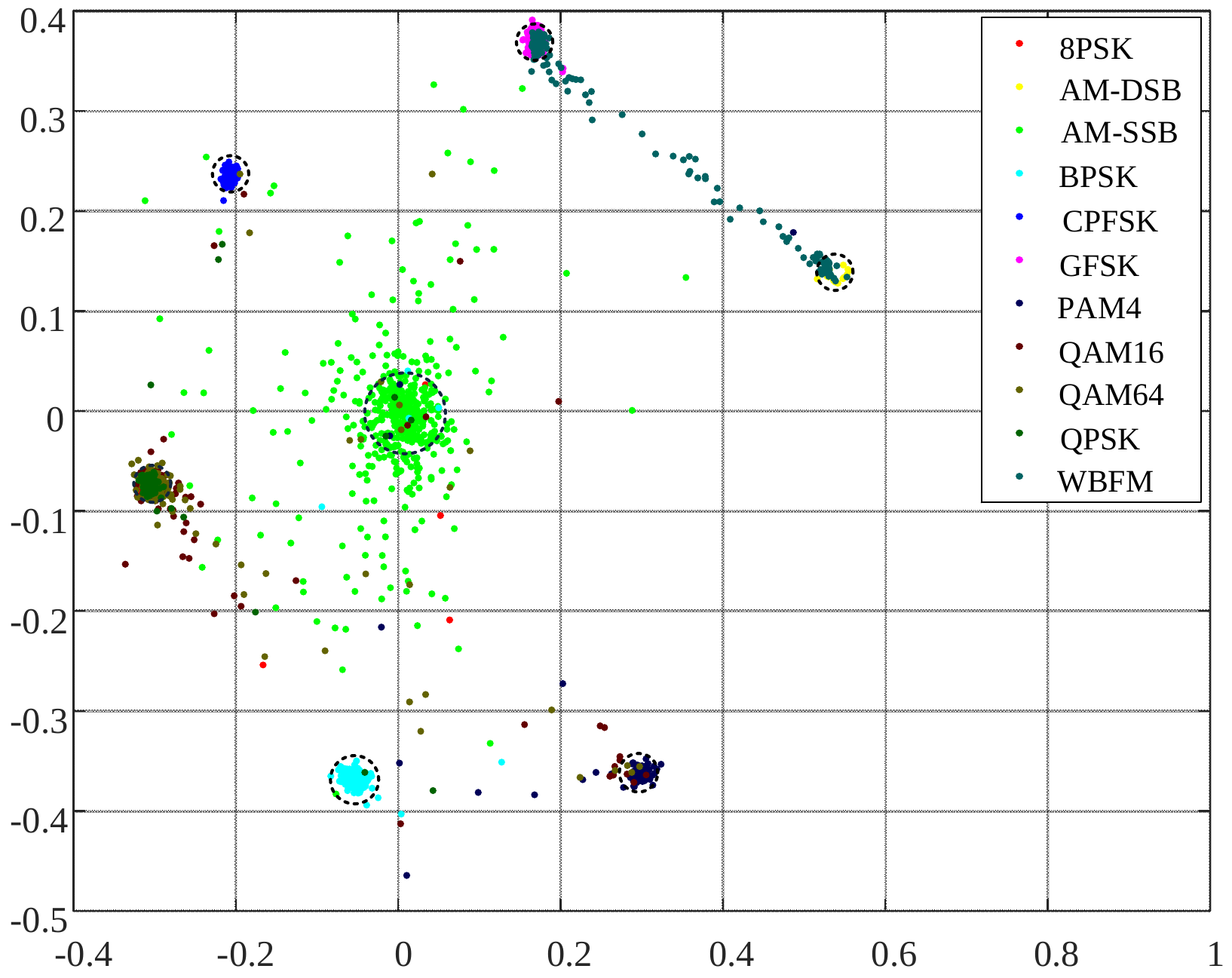}
  \caption{Features distribution of testing set.}
  \label{testft}
\end{figure}

{}\setlength{\parindent}{2em}The fully connnected layer with two neurons is added between concatenaion layer and output layer 
for visualization convenience. Modified model is retrained in the same way, the feature distributions of training set and testing 
set with CL-WF model are illustrated in Fig. \ref{trainft} and Fig. \ref{testft} respectively. The black dash circles indicate the 
known classes areas acquired from training, the other areas represent unknown classes or wrongly classified classes.%
\par{}\setlength{\parindent}{2em}From Fig. \ref{trainft} and Fig. \ref{testft} it can be concluded that: firstly, even open set 
unknown modulations would be clustering into a center; secondly, the centers of unknown classes are different from the known 
centers, this fact provides opportunity to recognize open set unknown classes. It should be awared that features are two-dimensional 
which caused extremely information compressed and dimension reducted, whereas this process is not adopted for performace evaluation. 
In addition, only correctly predicted training examples are used to calculate the centers of each class and the coefficients of 
different distributions.%


\section{Conclusion}
A new DC-LSTM model, center loss and weibull distribution based open set automatic modulation recognition algorithm is proposed 
in this letter. Two matrices are adopted to represent raw modulation signals, a loss function based on cross-entropy and center 
loss is used to extract separable and discriminative features which are used to fit the weibull distribution to recognize open 
set modulations. Proposed algorithm is validated on the public dataset. Quantitative analysis indicates that with no unknown 
classes, the DC-LSTM model could achieve an average accuracy of 90.2\% which achieves the state of the art result. With unknown 
classes, the performace can be dramatically improved on 14.2\%. Experiments results reveal that the proposed algorithm can 
effectively recognize modulation types even with unknown modulation types. This is the first effort to tackle the automatic 
modulation recognition under open set circumstance to our best knowledge.%

\ifCLASSOPTIONcaptionsoff
  \newpage
\fi

\bibliographystyle{IEEEtran}
\bibliography{IEEEabrv,DCLSTM}

\end{document}